\documentclass[twocolumn,showpacs,preprintnumbers,amsmath,amssymb]{revtex4-1}

\usepackage{graphicx}
\usepackage{amssymb}
\usepackage{hyperref}

\begin{document}

\title{From itinerant to local transformation and critical point in Ni-rich Ce$_2$(Ni$_{1-y}$Pd$_y$)$_2$Sn}

\author{J.G. Sereni $^1$, G. Schmerber $^2$, M. G\'omez Berisso $^1$ and J.P. Kappler $^2$}
\address{$^1$Low Temperature Div. CAB - CNEA, Conicet, 8400 Bariloche, Argentina}
\address{$^2$ IPCMS, UMR 7504 CNRS-ULP, 23 rue de
Loess, B.P. 43 Strasbourg Cedex 2, France}

\date{\today}


\begin{abstract}

Structural, magnetic ($M$) and thermal (C$_m$) studies on
Ce$_2$(Ni$_{1-y}$Pd$_y$)$_2$Sn alloys are presented within the $0
\leq y \leq 0.55$ range of concentration, showing evidences for
itinerant to local electronic transformation. At variance with
RKKY type interactions between localized moments $\mu_{eff}$, the
substitution of Ni by isoelectronic Pd leads the antiferromagnetic
transition to decrease from $T_N ~ 3.8$\,K to $~ 1.2$\,K between
$y=0$ and 0.48, while $M(H)$ measured at $H=5$\,T and 1.8\,K rises
from $0.12$ up to $0.75\mu_B$/Ce-at. Furthermore, the C$_m$(T$_N$)
jump increases with concentration whereas $|\theta_P|$ decreases.
The magnetic entropy $S_m(T)$ grows moderately with temperature
for $y=0$ due to a significant contribution of excited levels at
low energy, while at $y=0.5$ it shows a incipient plateau around
$S_m=Rln2$. All these features reflect the progressive ground
state transformation of from itinerant to a local character.

Another peculiarity of this system is the nearly constant value of
$C_m(T_N)$ that ends in an {\it entropy bottleneck} as $T_N$
decreases. Consequently, the system shows a critical point at
$y_{cr} \approx 0.48$ with signs of ferromagnetic behavior above
$H_{cr} \approx 0.3$T. A splitting of the $C_m(T_N)$ maximum,
tuned by field and concentration, indicates a competition between
two magnetic phases, with respective peaks at $T_N \approx 1.2$\,K
and $T_I \approx 1.45$\,K.

\end{abstract}


\maketitle

\section{Introduction}

Itinerant magnetism is mostly observed in U or Pu-based compounds
because of their $5f$ orbitals \cite{DeLong1} whereas a local
character is mostly expected in $4f$ Ce and Yb ones. However, one
basic question, not fully elucidated yet, concerns the possibility
of a local to itinerant transformation of the Ce-$4f^1$ orbital.
This possibility was investigated for 30 years within the so
called {\it local-itinerant dilemma} \cite{Mackintosh}. Although
the $4f^1$ delocalization is well documented among systems showing
$4f$ and conduction band hybridization \cite{sereni95}, that
mechanism implies the screening of the magnetic moment with the
consequent weakening of the magnetic order. In spite of that,
there are Ce compounds showing magnetic order with record high
values, being the outstanding examples CeRh$_3$B$_2$ (
$T_C=115$\,K \cite{Dhar}), CeScGe ($T_N=47$\,K \cite{CeScGe}) and
CeRh$_2$Si$_2$ ($T_N= 36$\,K \cite{CeRh2Si2}), which are claimed
to exhibit itinerant magnetism.

One characteristic of a well localized Ce-$4f$ orbital is related
to the fact that the doublet ($N=2$) ground state (GS) is not
significantly affected by Kondo screening (T$_K$) and the excited
crystal field (CEF) levels have a comparatively large splitting
($\Delta_{CEF}$), i.e. $T_K << \Delta_{CEF}$. The opposite
scenario, with $T_K \geq \Delta_{CEF}$, is the proper situation
where to search for itinerant magnetic behavior provided that the
magnetic order still occurs, which only occurs in a few system. As
a consequence of an eventual $N_{eff}>2$ character, the magnetic
contribution to the entropy ($S_m$) is expected to overcame the $R
Ln2$ value at relatively low temperature.

These conditions were recently observed in the ternary system
Ce$_{2}$(Pd$_{1-x}$Ni$_x$)$_2$Sn \cite{Pdrich}. There, the
substitution of Pd atoms by isoelectronic Ni leads to a change
from local (Pd-rich side) to itinerant (Ni-rich side) character.
Such a change in the electronic structure is associated to a
crystallographic modification from tetragonal Mo$_{2}$FeB$_{2}$ to
centered orthorhombic structure of W$_2$CoB$_2$ type
\cite{DiSalvo95} between $0.25 \leq y \leq 0.35$, where a gap of
miscibility occurs.

With the aim to compare the Ce-$4f$ orbital nature in these two
phases, the magnetic properties of the
Ce$_{2}$(Ni$_{1-y}$Pd$_y$)$_2$Sn alloys on the Ni-rich side (i.e.
within the orthorhombic structure) were investigated. Apart from
the usual magnetic characterization, for a proper analysis of the
itinerant to localized evolution of the system we focus on the
study of two basic parameters: the hight of the specific heat
anomaly at $T=T_N$ and the thermal dependence of the entropy
$S_m(T)$. These quantities allow to evaluate the role of the Kondo
effect and how the excited CEF levels contribute to the ground
state properties.

\section{Experimental results}

\subsection{Magnetic Susceptibility}

\begin{figure}[h]
\includegraphics[width=17pc]{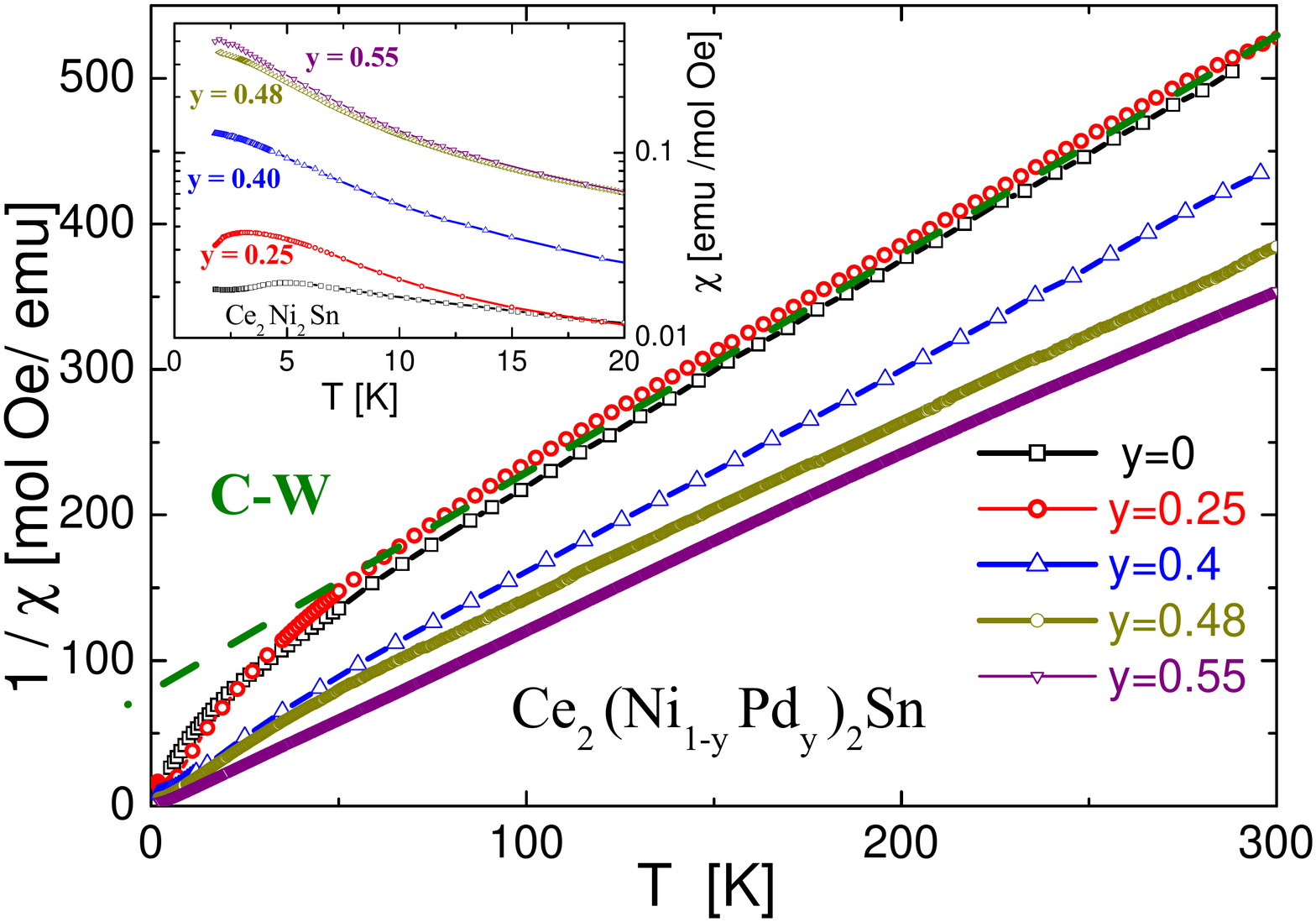}\hspace{4pc}%
\begin{minipage}[b]{14pc}\caption{\label{F1} High temperature inverse susceptibility
of Ni-rich alloys measured with $\mu_0 H=0.1$\,T. Inset: low
temperature magnetic susceptibility in a logarithmic scale.
\vspace{1pc}}
\end{minipage}
\end{figure}

The high temperature magnetic behavior ($T\geq 80$\,K) is well
described by a Curie-Weiss type contribution ($\chi =
C_C/(T-\theta_P)$) plus a minor Pauli contribution ($\chi_P = 5\pm
2 \times 10^{-3}$\,emu/mol Oe). In Fig.~\ref{F1} the inverse of
$\chi$ is presented to extract the respective paramagnetic
temperatures $\theta_P(y)$ and effective moments $\mu_{eff}(y)
\propto \surd C_C$. The $\theta_P(y)$ values range between -58\,K
at the Ni-rich limit and $\approx 6$ at $y=0.55$. Simultaneously,
$\mu_{eff}(y)$ increases from $\mu_{eff}=2.3\mu_B$ up to $\approx
2.54\mu_B$ between those concentration values. A detail of the
$\chi(T)$ dependencies at low temperature ($T<20$\,K) are included
in the inset of Fig.~\ref{F1} using a logarithmic scale to cover
the large variation of $\chi(y)$ at $T=1.8$\,K.

\begin{figure}[h]
\includegraphics[width=15pc]{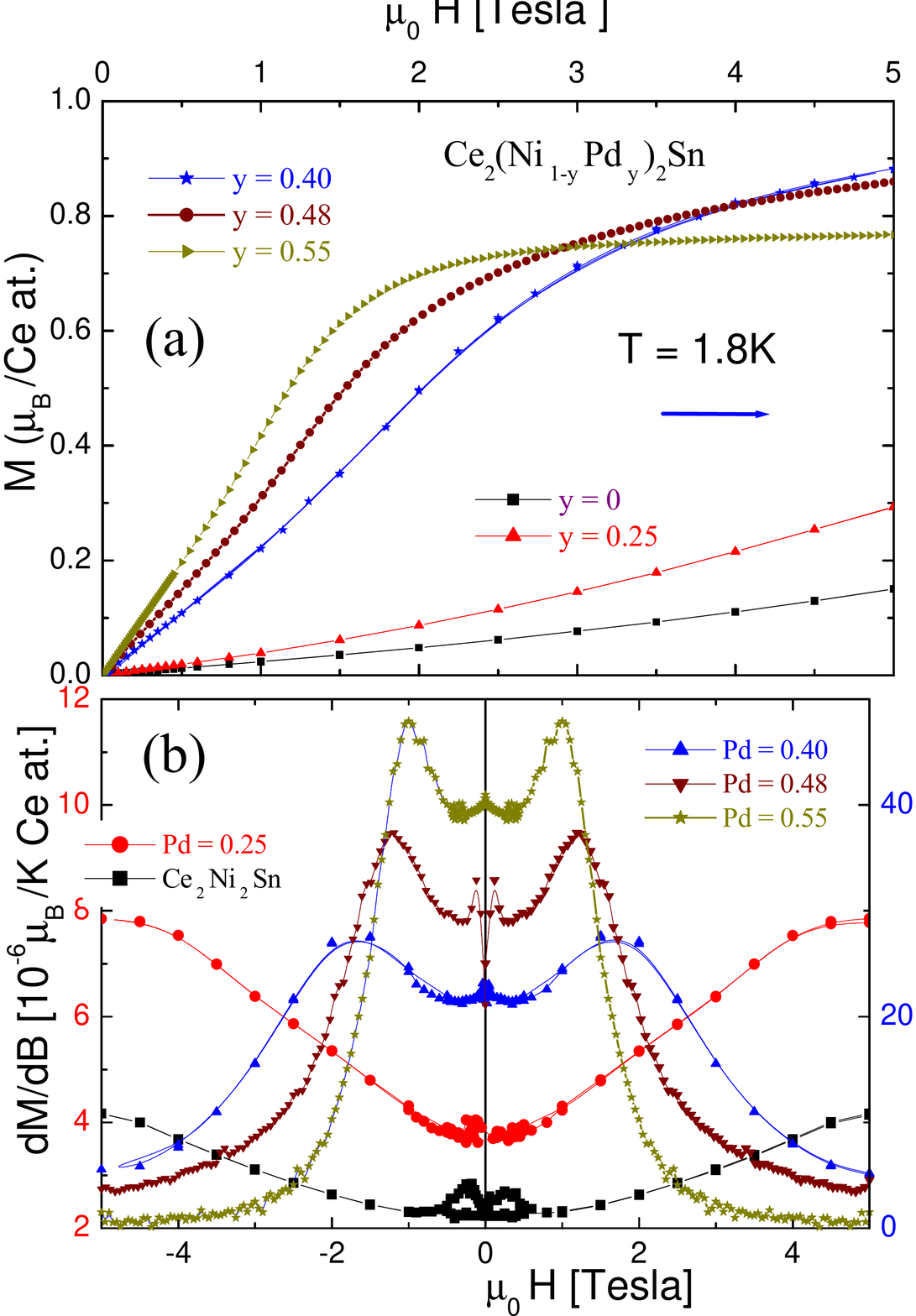}\hspace{4pc}%
\begin{minipage}[b]{14pc}\caption{\label{F2} a) Field dependent magnetization of the
1.8\,K isotherms for the Ni-rich range measured up to 5T. b) dM/dH
vs H derivative showing the difference between Ni-rich samples
(left axis) and those with similar Pd/Ni concentration (right
axis). \vspace{1pc}}
\end{minipage}
\end{figure}

In order to explore this strong increase of $\chi(y)$ at low
temperature the isothermal field dependence of the magnetization
$M(H)$ was measured at $T=$1.8\,K, 3\,K and 4\,K up to $\mu_0
H=5$\,Tesla. A clear difference in the $M(H)$ dependence can be
appreciated in the results obtained at $T=1.8$\,K between Ni-rich
samples and those with similar Pd/Ni concentration (see
Fig.~\ref{F2}a). While $M(H)$ shows a slight positive curvature
with low values of $M$ in $y=0$ ($0.15\mu_B$) and 0.25 alloys,
those with $y \geq 0.4$ show a tendency to saturation ($M_{sat}$)
at $H=5$Tesla. Despite the fact that the former alloys have a $T_N
\geq 1.8$\,K while in the latter $T_N < 1.8$\,K, similar
measurements performed at $T=4$\,K (i.e. in the paramagnetic
phase) show the same variation. Notably, $M_{sat}(y)$ undergoes a
maximum of $0.88\mu_B$ at $y=0.40$ and then slightly decreases to
$0.76\mu_B$ for $y=0.55$. In coincidence with the $M_{sat}$
decrease at these concentrations, the initial slope of $M(H)$
increases as a sign of a stronger interaction between magnetic
moments. Additionally, a further slight increase in the slope is
observed around $H=1$\,Tesla. To remark these features, the
$\partial M/\partial H$ derivative is presented in Fig.~\ref{F2}b,
showing that the change of slope only occurs in the alloys with
strong magnetization (i.e. $y\geq 0.4$) and it grows up to
$y=0.55$. Notably, the $M(H)$ dependence above that critical field
is properly described by a $\tanh(H)$ function which characterizes
ferromagnetic systems.

\begin{figure}[h]
\includegraphics[width=15pc]{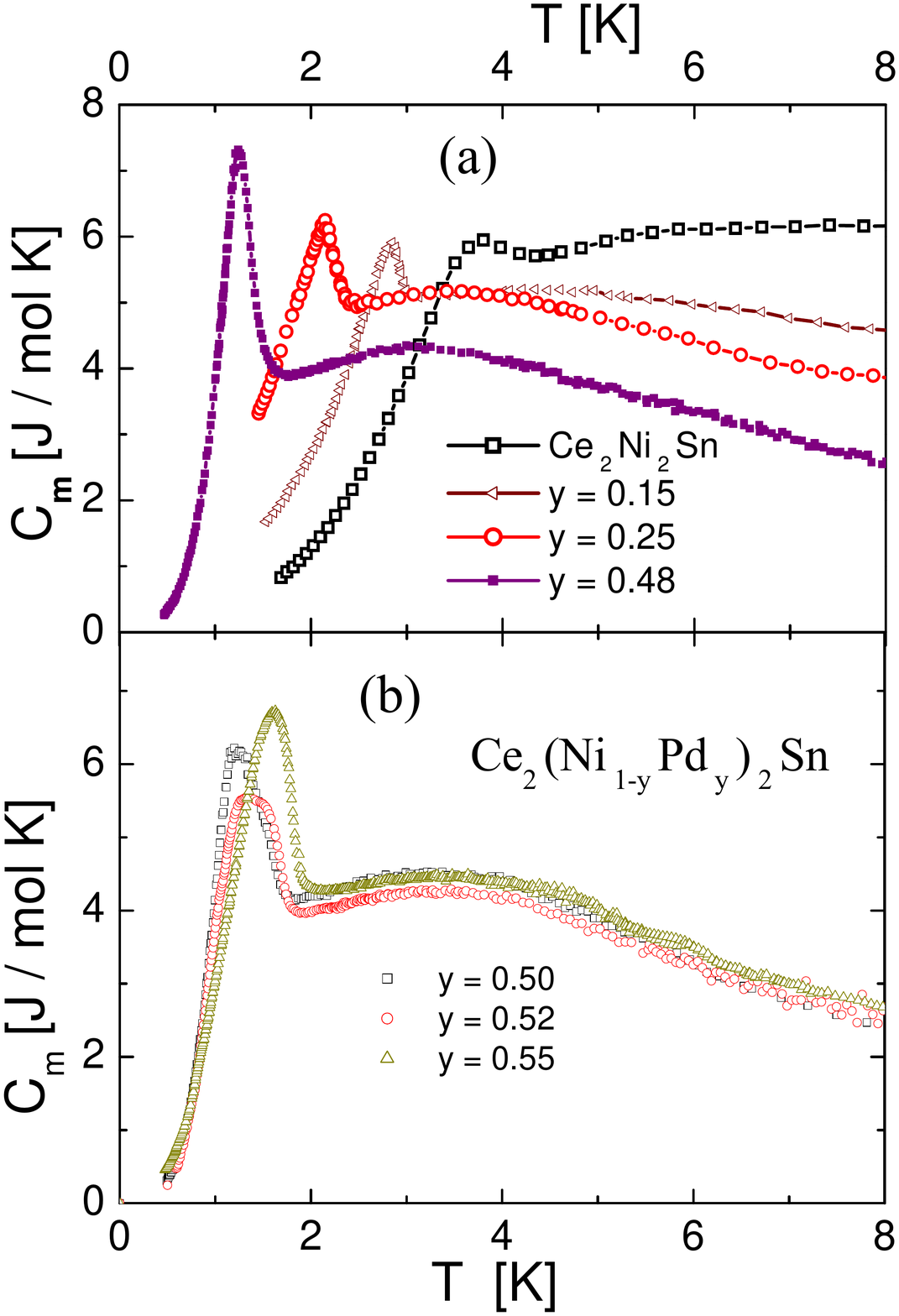}\hspace{4pc}%
\begin{minipage}[b]{14pc}\caption{\label{F3} Thermal dependence of the magnetic
contribution to the specific heat $C_m$ for: a) Ni-rich alloys and
b) critical region at similar Ni/Pd concentration. \vspace{1pc}}
\end{minipage}
\end{figure}

\subsection{Specific heat}

The magnetic contribution ($C_m$) to the measured ($C_P$) specific
heat was obtained by subtracting the phonon contribution extracted
from the reference isotypic compound La$_2$Ni$_2$Sn as $C_m = C_P
- C_{La}$. From the thermal dependence of $C_m$ one can see in
Fig.~\ref{F3}a that the ordering temperature $T_N(y)$, defined as
the temperature of the maximum of $C_m(T)$, decreases with Pd
concentration. This evolution extrapolates to $T_N=0$ at $y
\approx 0.70$, beyond the structural stability of this system.
However, between $0.48\leq y \leq 0.52$, $T_N(y)$ remains nearly
constant at $T_N \approx 1.2$\,K while another anomaly arises for
$y=0.55$ at $T_I \approx 1.5$\,K as shown in Fig.~\ref{F3}b.

\begin{figure}[h]
\includegraphics[width=15pc]{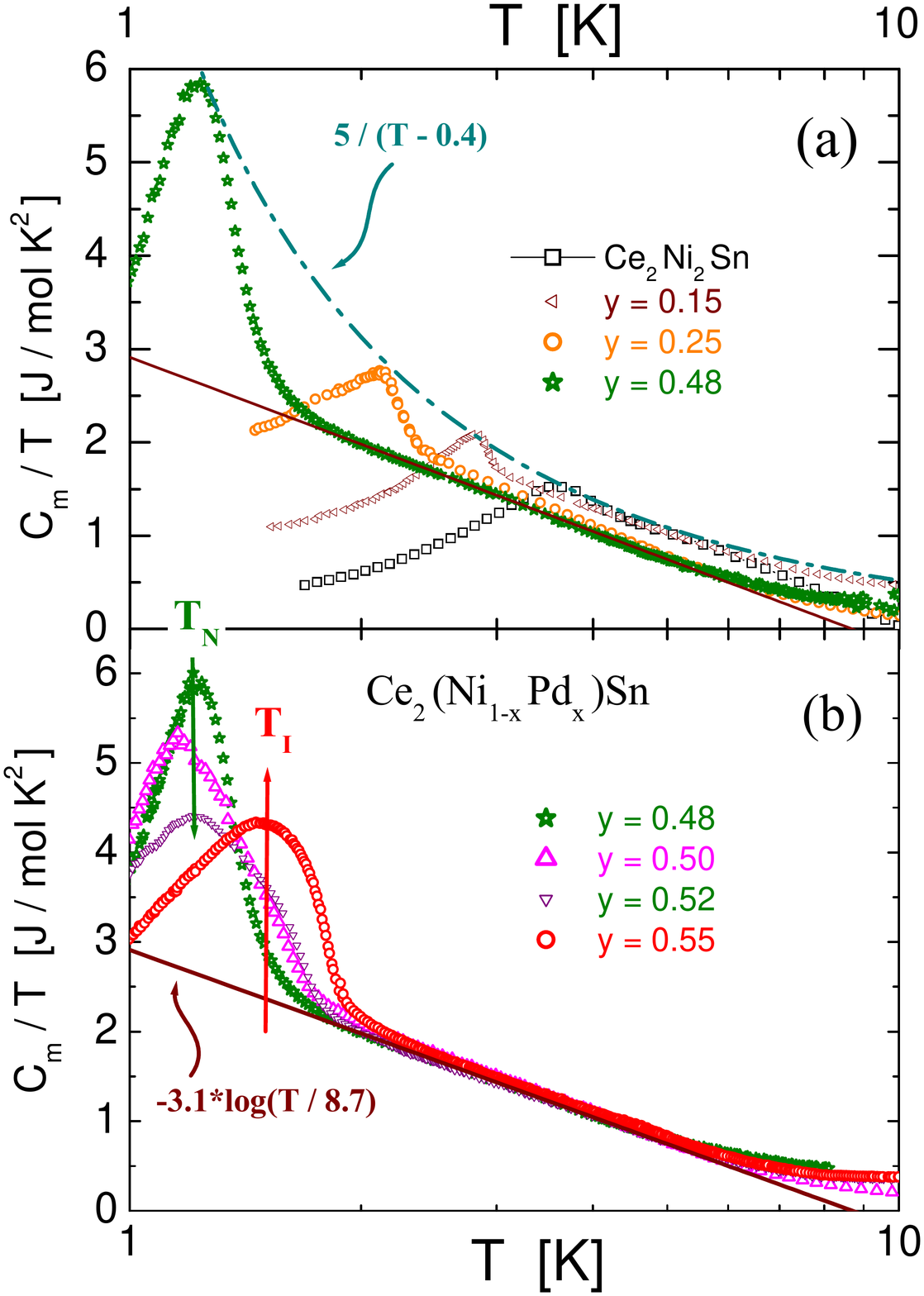}\hspace{4pc}%
\begin{minipage}[b]{14pc}\caption{\label{F4} Logarithmic dependence of the magnetic
contribution to specific heat divided by temperature $C_m/T$ for:
(a) $0 \geq y \geq 0.48$ samples. Dash-dot curve describes the
$C_m/T(y)$ maxima dependence. (b) Alloys with Pd/Ni equi-atomic
concentrations. Arrows at $T_N$ and $T_I$ indicate a two phases
competition. Straight line represents a logarithmic reference
fitting a universal function (see the text). \vspace{1pc}}
\end{minipage}
\end{figure}

While the $C_m(T)$ dependence at $T<T_N$ is typical for an
antiferromagnetic system, the $C_m(T)$ maximum ($C_{max}$) behaves
quite unusual because its value slightly changes up to $y = 0.48$
and becomes sharper as Pd concentration increases, see
Fig.~\ref{F3}a. According to the law of corresponding states,
$C_{max}$ should decrease with decreasing $T_N(y)$, to become zero
at $T=0$. As a consequence the magnetic transition with chances to
reach zero temperature have to show $C_{max}/T_N \approx$ const.
\cite{JLTP}. On the contrary, this system shows $C_{max}/T$ rising
as $\propto 1/T_N(y)$ as indicated in Fig.~\ref{F4}a. This
divergent $C_{max}/T_N$ dependence places the $y=0.48$ alloy at a
critical concentration point.

Above $T_N(y)$, $C_m(T)$ shows a very broad maximum which
progressively decreases in temperature and intensity as Pd
concentration increases, see Fig.~\ref{F3}. Such a temperature
dependence is characteristic of non-Fermi-liquids (NFL)
\cite{HvL,Stewart} in the region of quantum critical behavior that
affects the thermal properties because of emerging quantum
fluctuations. These low energy fluctuations overcome classical
thermal fluctuations below 2 or 3\,K as it was observed in Ce NFL
systems \cite{PhilMag}. Within that regime, a characteristic
$C_m(T)/T = - A \log(T/T^*)$ thermal dependencies occurs, like the
observed in this system above $T_N$, see Fig.~\ref{F4}. The
parameter $T^*$ represents a characteristic energy scale that
describes the range of the $C_m(T)/T$ tail. The obtained $A$ and
$T^*$ parameters allow to scale this system into an universal
curve $C_m/t = -7.2 \log(t)$, with $t=T/T^*$, extracted for a
number of Ce intermetallic compounds showing logarithmic thermal
dependencies \cite{scaling}.

\section{Discussion}

\subsection{Itinerant to local electronic transformation}

Comparing magnetic and thermal properties one observes that,
within the $0 \leq y \leq 0.48$ range, this system presents an
unusual discrepancy between the decreasing $T_N(y)$ and the
increasing amplitude of the $\Delta C_m(T_N)$ jump, see
Fig.~\ref{F3}a. The latter behavior is in line with the increasing
values of magnetization presented in Fig.~\ref{F2}a. In an usual
scenario of competing RKKY magnetic interactions with Kondo
effect, the former mechanism tends to increase magnetic order
whereas the later to screen magnetic moments \cite{Lavagna}. This
classical description does not apply to this system because a
$T_N(y)$ decrease would imply a weakening of the RKKY interaction
due to an enhancement of the Kondo screening and the consequent
decrease of $\Delta C_m(T_N)$ \cite{Pdrich}. On the contrary, the
observed growing and narrowing of $\Delta C_m(T_N)$ suggests a
tendency towards the localization of Ce-$4f$ states driven by a
weakening of $4f$-band hybridization (i.e. a decrease of
$T_K(y)$). Furthermore, the ferromagnetic type dependence of
$M(H)$ above $H_{cr} \approx 0.5$\,Tesla observed in the $y\geq
0.48$ alloys clearly suggests a minor role of Kondo effect at this
range of concentration. These facts make evident that different
mechanisms are involved in the evolution of the ground state
properties of this system.

The most likely is a decreasing contribution of the excited
crystal electric field CEF levels at low energy, which leads the
system into an itinerant to localized transformation. Taking into
account that $|\theta_P|$ can be considered $\propto T_K$
\cite{Krishna}, if $|\theta_P|$ is comparable to the splitting
$\Delta_{CEF}$ of the first CEF level the contribution of that
doublet is relevant, but it becomes irrelevant once $|\theta_P| <<
\Delta_{CEF}$. The reported decrease of $|\theta_P|(y)$ from 58\,K
(at $y=0$) to 6\,K (at y=0.55), together with the increase of
$\mu_{eff}$ and $M(H)$ with concentration, support this
description. From $C_m(T)$ measurements up to $T=30$\,K on sample
$y=0.55$, the first CEF splitting was evaluated as $\Delta_{CEF} =
40\pm 5$\,K.

\begin{figure}[h]
\includegraphics[width=17pc]{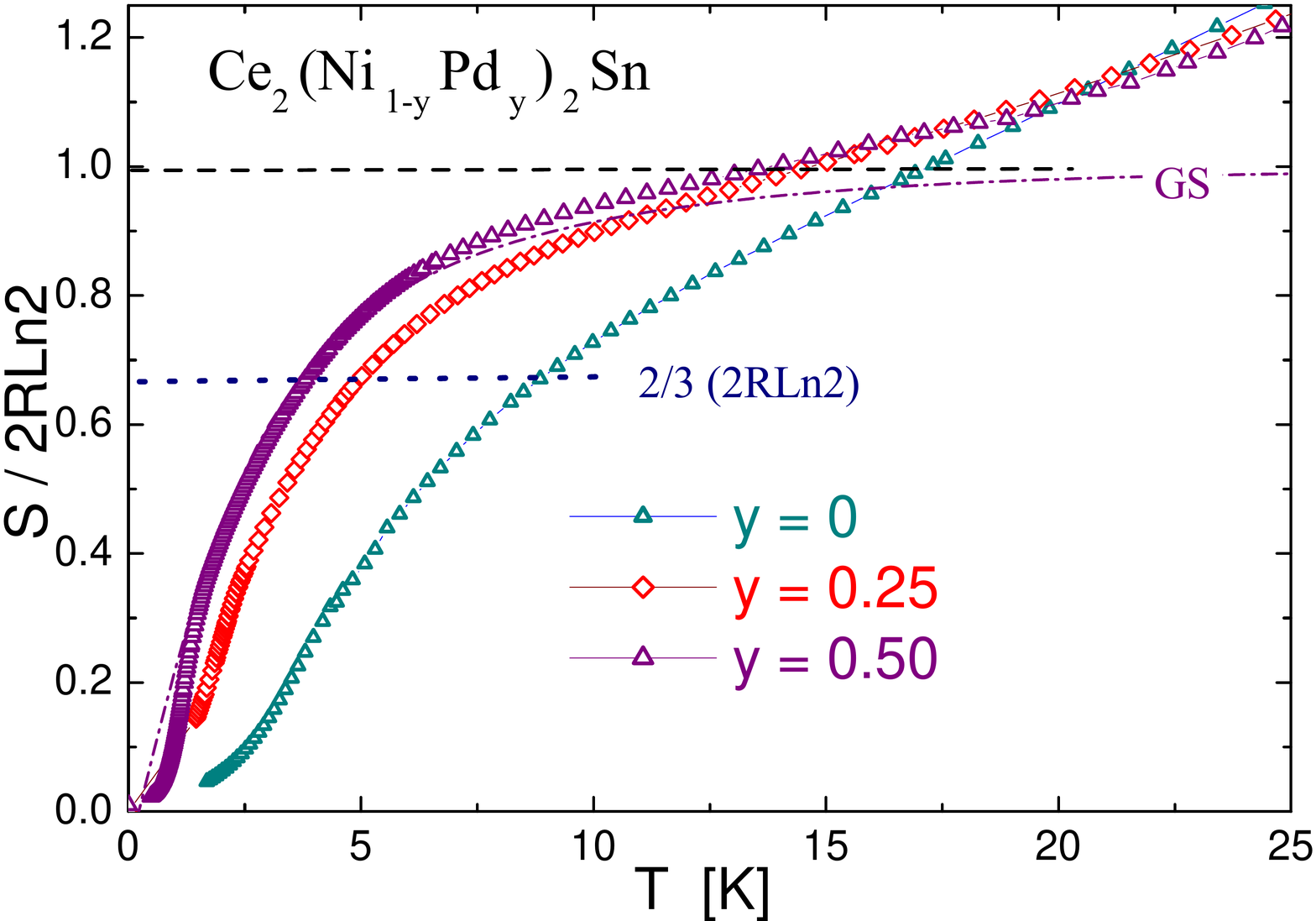}\hspace{2pc}%
\begin{minipage}[b]{14pc}\caption{\label{F5} Thermal dependence of the entropy (normalized to one Ce at.) for
$y=0$ and the higher Pd concentrations. Dashed curve is a
qualitative representation for the ground state contribution in
sample $y=0.55$.}
\end{minipage}
\end{figure}

An alternative way to test this scenario is the analysis of the
entropy variation for different concentrations, $S_m(y,T)$. In
Fig.~\ref{F5} we show the thermal evolution of the magnetic
entropy computed as $S_{m}=\int C_m/T \times dT$. On the Ni-rich
limit, one sees that Ce$_2$Ni$_2$Sn presents a moderate and
monotonous increase of $S_m(T)$ that overcomes $S_m=2RLn2$ around
15\,K. This behavior reveals a significant broadening of ground
and excited CEF levels that partially overlap each other because
of comparable $T_K$ and $\Delta_{CEF}$ energies. On the contrary,
the $y=0.50$ and 0.55 alloys show an incipient plateau of $S_m(T)$
around $T\geq 10$\,K. This change confirms that the doublet GS is
mostly occupied at that temperature whereas the first excited CFE
doublet has reduced significantly its contribution to the GS
properties. Applying Desgranges-Schotte \cite{Desgr} criterion of
$S_{m}(T_K) \simeq 2/3 R\ln2$ for single Kondo moments, the
extracted values of $T_K(y)$ clearly show a decrease of Ce-$4f$
orbitals hybridization. This is the key to understand the unusual
behavior of this system that can be explained as a combination
between the decrease of the {\it 4f-band} hybridization intensity
and the consequent reduction of the CFE levels contribution to the
low energy density of states.

\subsection{Critical region and Phase Diagram}

Another peculiarity of this system emerges from the analysis of
the $S_m(T)$ dependence at low temperature since the entropy gain
at $T_N$ is remarkably low, i.e. $\approx 0.3$\,Rln2 per Ce-at.
Such a low value is observed in Ce and Yb compounds which do not
order because of e.g. magnetic frustration effects
\cite{bottleneck} but present a $C_m(T)/T$ anomaly characterized
by a divergent power law tail above the temperature of the maximum
$T_{max}$. Although in Ce$_2$(Pd$_{1-y}$Ni$_y$)$_2$Sn the tail
follows a $- \log(T/T^*)$ dependence, the $C_m/T$ maximum diverges
at $T_N(y)$ following a: $5/(T-0.4)$\,K$^{-1}$ dependence, as
depicted in Fig.~\ref{F4}a, that should end in a critical point
because of an {\it entropy bottleneck} \cite{PhilMag}. This type
of behavior was also observed in URu$_2$Si$_2$ under magnetic
field \cite{Jaime} and CeTi$_{y}$Sc$_{1-y}$Ge \cite{CeTiScGe}. The
latter shows a very similar divergency of its $C_m/T$ maximum also
approaching a structural instability.

\begin{figure}[h]
\includegraphics[width=18pc]{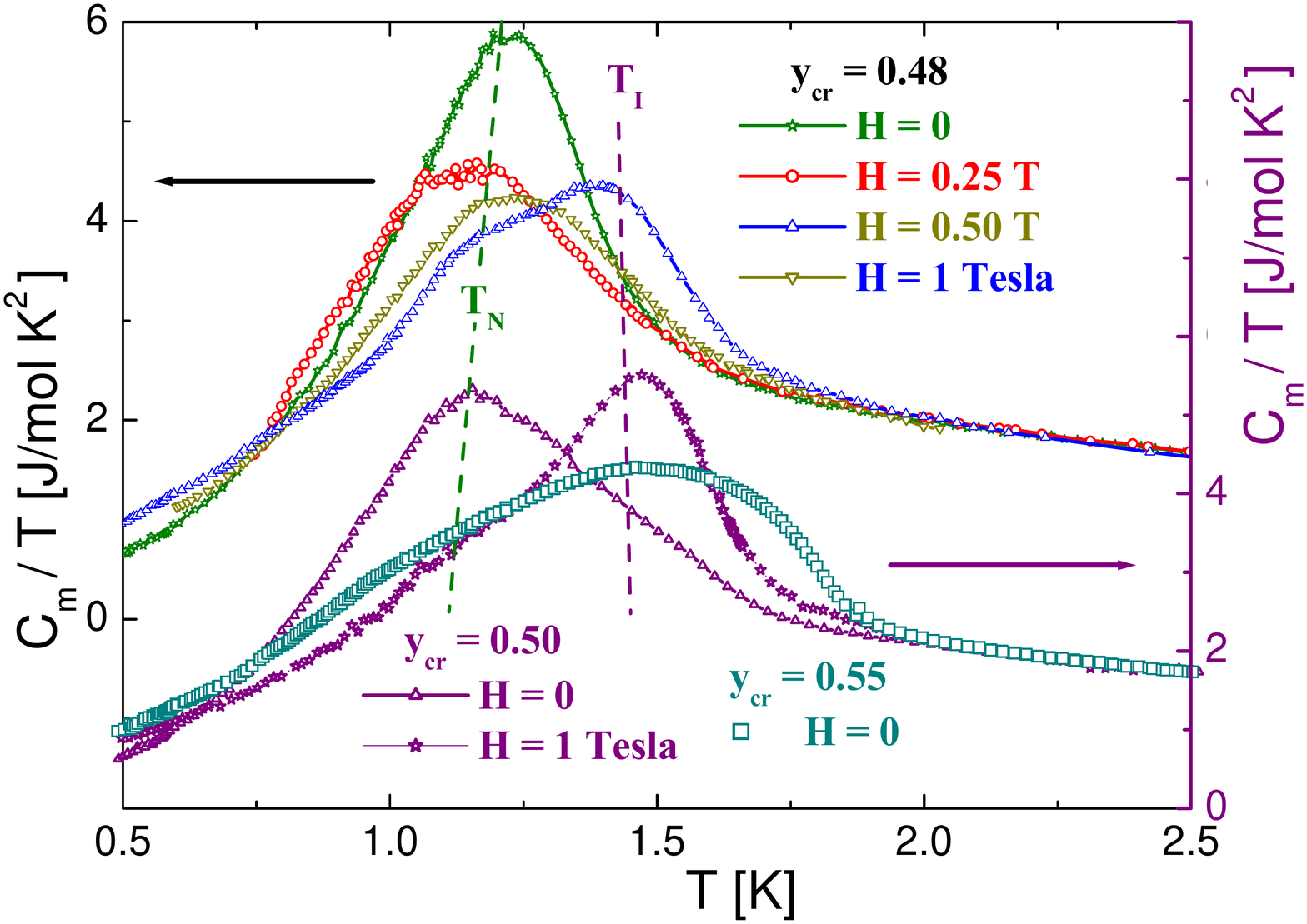}\hspace{4pc}%
\begin{minipage}[b]{14pc}\caption{\label{F6} Left axis (shifted up): specific heat of the
magnetic anomaly of sample $y_{cr}=0.48$ at zero and three
magnetic fields. Right axis: Comparison with other two
concentrations. Dashed lines mark the mean temperatures of the
anomalies at $T_N \approx 1.2$\,K and $T_I \approx 1.5$\,K.
\vspace{1pc}}
\end{minipage}
\end{figure}

In Fig.~\ref{F4}a and b one can see how the temperature of the
magnetic anomaly decreases down to $T_N \approx 1.2$\,K at $y_{cr}
= 0.50$ where another contribution at $T_I \approx 1.45$\,K arises
while the intensity of the former decreases. A detailed study of
those specific heat anomalies around the critical region is
presented in Fig.~\ref{F6}, performed at zero and applied magnetic
field. Under magnetic field, the anomaly of the $y_{cr}=0.48$
alloy splits into two maxima. In that figure one can clearly see
how the $C_m/T$ maximum at $T_N \approx 1.2$\,K decreases in
intensity while the one at $T_I$ increases. The intrinsic
character of this competition between a vanishing phase and an
emerging one is confirmed by the same behavior observed in the
$y=0.50$ alloy, right axis in Fig.~\ref{F6}. Beyond that
concentration, sample $y=0.55$ shows how the relative
contributions are reversed. The growing intensity of the $T_I$
anomaly with field and the $M(H)$ dependence close to that
temperature make evident a ferromagnetic character of this alloy
despite of the lack of hysteresis in the $M(H)$ curves.

\begin{figure}[h]
\includegraphics[width=18pc]{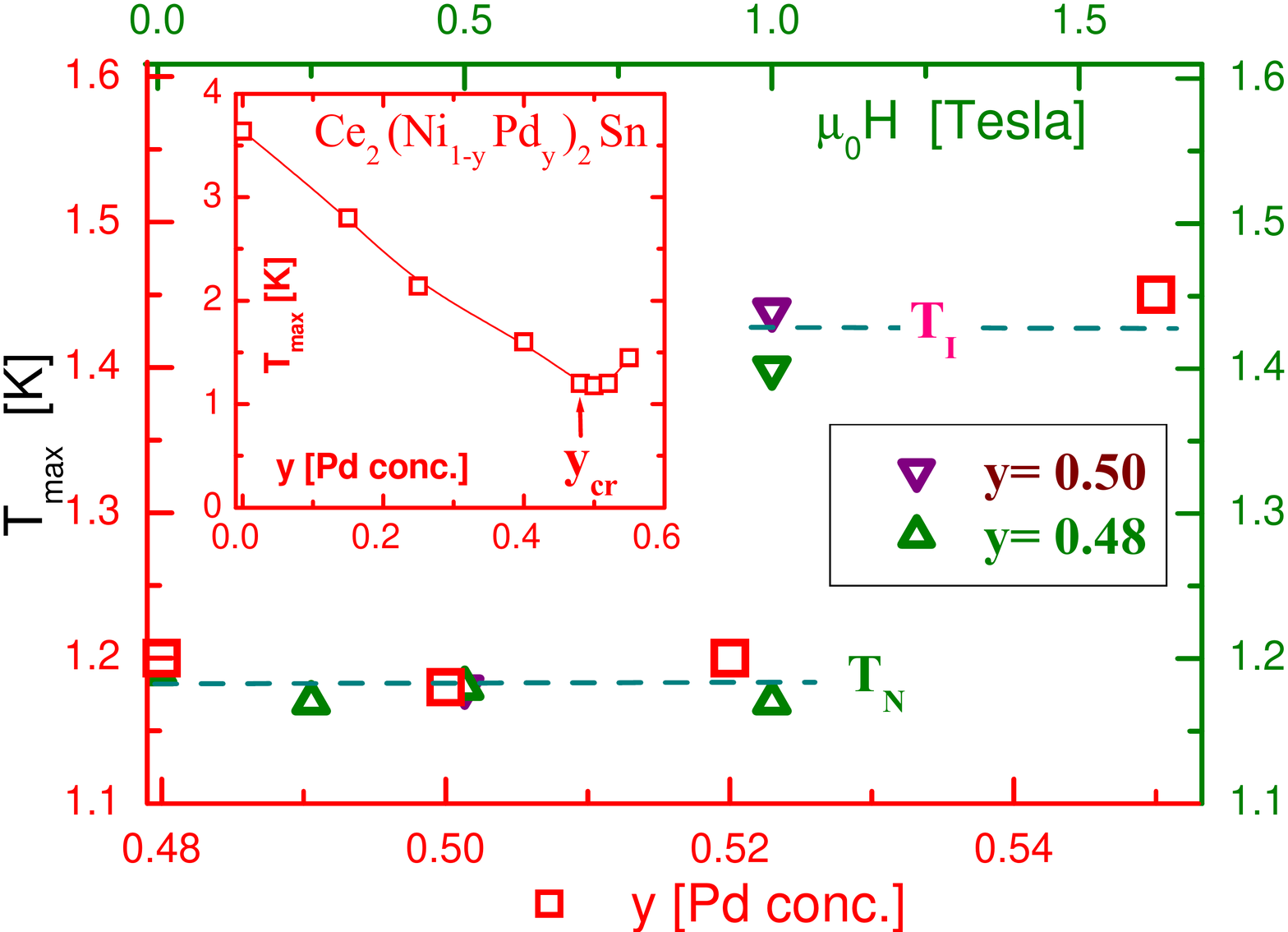}\hspace{2pc}%
\begin{minipage}[b]{14pc}\caption{\label{F7} \small Magnetic phase
diagram around the critical point comparing $T_N(0.48\leq y \leq
0.55)$ ($\square$) and field dependencies of $y_{cr}=0.48$
($\triangle$) and $y=0.50$ ($\triangledown$). Inset: full Pd
concentration range for $T_N(y)$.}
\end{minipage}
\end{figure}

The overall concentration dependence of $T_N(y)$ is included in
the inset of Fig.~\ref{F7}, showing its decrease from 3.8\,K down
to 1.2\,K at $y_{cr}=0.48$. At that concentration the competition
between two phases occurs as it shown by specific heat
measurements presented in Fig.~\ref{F6}. Those details are
summarized in the main body of Fig.~\ref{F7}. There, concentration
and magnetic field effects on both contributions with respective
maxima at $T_N\approx 1.2$\,K and $T_I\approx 1.45$\,K are
compared. Notice that thermal ($k_B T_{max}$) and magnetic ($\mu_B
H$) energy scales used in the phase diagram are comparable.

\section{Conclusions}

The outstanding characteristics observed in the
Ce$_2$(Ni$_{1-y}$Pd$_y$)$_2$Sn family are the transformation of
the magnetic properties from itinerant to mostly localized and the
presence of a critical concentration with signatures of
ferromagnetic contribution.

The former characteristic is recognized from the Pd concentration
dependence of the magnetic properties like: i) the decrease of
$|\theta_P|$, reflecting a decrease of $T_K$, ii) the increase of
the effective magnetic moment and iii) the sharpening of the
specific heat transition. All these properties behave at variance
to the decrease of the ordering temperature $T_N$ in a usual RKKY
scenario of localized moments. In this system, the $T_N(y)$
decrease is driven by the reduction of the CEF excited levels
contribution to the ground state properties as revealed by the
thermal evolution of the magnetic entropy with growing Pd
concentration.

The second peculiarity of this system concerns the fact that the
$C_m(T)$ anomaly associated to $T_N(y)$ shows a nearly constant
maximum which ends into a {\it bottleneck} of the magnetic entropy
at a critical concentration $y_{cr}\approx 0.48$. At that point
$M(H)$ reveals a ferromagnetic type dependence above $H_{cr}
\approx 0.4$T. The splitting of the $C_m(T)$ maximum, tuned by
field and concentration indicates a competition between two
magnetic phases within the $0.48\leq y \leq 0.55$ range. Their
respective characteristic energies are $T_N \approx 1.2$\,K and
$T_I \approx 1.45$\,K, with the former contribution vanishing and
the latter arising as a function of concentration and/or field.

Lower temperature magnetic measurements (i.e. $T<1.8$\,K) are
required to better characterize the magnetic ground state within
the critical region. Spectroscopic measurements may shed light
into the dynamical nature of these competing phases and band
calculations are highly desirable to trace the modification of the
Fermi surface.

\end{document}